\documentclass[a4paper]{article}

\usepackage{INTERSPEECH2019}
\usepackage{balance}

\PassOptionsToPackage{hyphens}{url}
\usepackage[bookmarks=false,colorlinks=false,hidelinks]{hyperref}
\usepackage{todonotes}

\usepackage[final]{changes}

\title{The GDPR \& Speech Data: Reflections of Legal and Technology Communities,\\ First Steps towards a Common Understanding}
\name{Andreas Nautsch$^1$, Catherine Jasserand$^2$, Els Kindt$^3$, \\ Massimiliano Todisco$^1$,  Isabel Trancoso$^4$ and
Nicholas Evans$^1$}
\address{
  $^1$Audio Security and Privacy, Digital Security Department, EURECOM, France\\
  $^2$Security, Technology \& e-Privacy Research Group,  University of Groningen, Netherlands\\
  $^3$Centre for IT \& IP Law (CITIP), KU Leuven, Belgium; eLaw, Universiteit Leiden, Netherlands\\
  $^4$INESC-ID / IST, University of Lisbon, Portugal
 }
\email{\{nautsch,todisco,evans\}eurecom.fr, c.a.jasserand@step-rug.nl, \\ els.kindt@kuleuven.be, isabel.trancoso@inesc-id.pt}

\begin{document}

\maketitle
\begin{abstract}
    Privacy preservation and the protection of speech data is in high demand, not least as a result of recent regulation, e.g.\ the General Data Protection Regulation (GDPR) in the EU. While there has been a period with which to prepare for its implementation, its implications for speech data is poorly understood. This assertion applies to both the legal and technology communities, and is hardly surprising since there is no universal definition of `privacy', let alone a clear understanding of when or how the GDPR applies to the capture, storage and processing of speech data.  In aiming to initiate the discussion that is needed to establish a level of harmonisation that is thus far lacking, this contribution presents some reflections of both legal and technology communities on the implications of the GDPR as regards speech data. The article outlines the need for taxonomies at the intersection of speech technology and data privacy---a discussion that is still very much in its infancy---and describes the ways to safeguards and priorities for future research.  In being agnostic to any specific application, the treatment should be of interest to the speech communication community at large.

\end{abstract}



\section{Introduction}

On the surface, the concept of privacy may appear to be quite straightforward.  In reality, however, the very notion of privacy is as challenging to define as the diversity of speech applications where there is potential for privacy intrusion. Intrusions into privacy can stem from the misuse of speech data for purposes other than those to which permission may have been granted or the processing or storage of speech data that may have been captured without consent.  Given the diversity and ubiquity of applications that now capture, store and process speech signals, the concept of privacy is indeed one that is difficult to define.


It is therefore perhaps not too surprising that privacy has no formal, legal definition.  Even so, regulation such as the General Data Protection Regulation (GDPR)~\cite{EU-Regulation679-2016} within the European Union implies certain restrictions and safeguards upon the use of speech data.  This situation is somewhat troubling since the legal and technical communities do not yet share a common understanding of what the existing regulation implies in terms of speech data and speech technology, and of how the existing technology is perceived and understood by legislation. The provision on and interpretation of the law depends on experts.

A common understanding will take time to evolve.  \replaced{This paper is a first attempt at establishing a common understanding. As the topic is complex, there is a need to pursue the interdisciplinary discussions to establish a level of harmonisation.}{This article aims to provide a starting point and to facilitate the discussions that will be needed in future in order to establish a level of harmonisation.}  While presenting some reflections from both the legal and technical communities, it is not intended to be an exhaustive treatment. Instead, it gently introduces some of the core issues and implications of privacy and data protection regulation upon the speech communication technical community and vice versa. 
A set of taxonomies is proposed intended as a basis for future dialogue between our two communities. 




This paper is organised as follows. Section~\ref{sec:legal} presents a legal perspective on the European privacy regulation.  Section~\ref{sec:tech} presents a technical perspective aimed at non-experts. A set of seven taxonomies are proposed in Section~\ref{sec:taxonomoies}, whereas Section~\ref{sec:conclusion} presents some conclusions.
\section{Privacy, a Legal Perspective}
\label{sec:legal}
This section provides guidance concerning the interpretation of privacy, what should be considered as biometric data as concerns relevant regulation, when data should be treated as being sensitive and the grounds for processing sensitive data.



\vspace{-0.25em}
\subsection{What are `privacy' and `data protection'?}
\vspace{-0.25em}

Even though privacy is a fundamental and enforceable right in generally all western democracies, it lacks a universal definition, even in legal provisions or in the courts. 
Privacy was originally defined in the US by Warren and Brandeis~\cite{WarrenBrandeis-RightToPrivacy-Harvard-1980} as `the right to be let alone'.  This right is viewed as `the foundation of individual freedom'~\cite{Gallaghar-2002}.  US scholars usually distinguish four types of privacy: informational privacy (also known as data privacy); physical privacy; decisional privacy, and proprietary privacy~\cite{Allen-UnderstandingPrivacyBasics-PractisingLawInstitute-2006}.
In the EU, the right to respect for privacy has no legal definition and `is a broad term not susceptible to exhaustive definition'. Despite this, the right to the respect for privacy is referenced explicitly in Art. 8 of the European Convention on Human Rights\footnote{Adopted in 1950 by the Council of Europe.} as well as in the Catalogue of Fundamental Rights and Freedoms in the EU in Art. 7 of the EU Charter. 

The contour (delimitation) of the right to privacy is, however, defined by case law of the European Court of Human Rights. As interpreted, private life is not restricted to the notion of an ‘inner circle’, but extends to various aspects relating to personal identity, including the right to develop relationships with others. The right encompasses, for instance, the protection of an individual's reputation, the protection of information about his/her health~\cite{ECHR-ZVSFinland-1997}, the right to personal development and autonomy, as well as the right to the protection of his/her personal data. Whether there is a risk of infringement or if the right has actually been infringed, shall therefore be reviewed case-to-case, and may become increasingly challenging in digital environments. 

Extracting or processing such information without safeguards could possibly infringe upon the privacy of the individual concerned, from case to case; the concept of privacy is, moreover, interpreted by each EU nation differently within their specification of the GDPR opening clauses.
At the same time, the use of speech data and extraction of additional information will also fall under personal data protection.  Hence, while the recording, processing and use of speech data may or may not violate privacy, the same will also fall under \emph{data protection} regulation, which is distinct from \emph{privacy} regulation.  

Both matters are strictly different, and while the fundamental right to respect for privacy is not defined, data protection regulation in the EU consists of both a fundamental right (Art. 8 in the EU Charter) and specific, detailed rules which need to be respected. This also applies to the use of speech in a research environment, although exceptions to particular obligations exist, as long as the fundamental rights are respected.  


\vspace{-0.25em}
\subsection{When does data qualify as `biometric data'?}
\vspace{-0.25em}
The GDPR introduces a new category of personal data: biometric data. These data are defined as `personal data resulting from specific technical processing relating to the physical, physiological or behavioural characteristics of a natural person, which allow or confirm the unique identification of that natural person, such as facial images or dactyloscopic data'~\cite[Art. 4(14)]{EU-Regulation679-2016}.

\vspace{-0.25em}
\subsection{When is data `sensitive'?}
\vspace{-0.25em}

The GDPR protects the processing of `sensitive' data (term used by legal experts), which is described as including `personal' data (term used by the GDPR) revealing racial or ethnic origin, political opinions, religious or philosophical beliefs' and the processing of `biometric data for the purpose of uniquely identifying a natural person, data concerning health or data  concerning [\dots{}] sex life or sexual orientation'~\cite[Art. 9(1)]{EU-Regulation679-2016}. The processing of such data is prohibited, except when its processing would be allowed under one of ten exceptions laid down in the GDPR---or under an exception specified in a legislation implementation of GDPR opening clauses by an EU nation. 

The GDPR requires the entity responsible for the processing (also known as the controller) to make a detailed assessment in case processing is `likely to result in a high risk to the rights and freedoms of natural persons'~\cite[Art. 35]{EU-Regulation679-2016}.  At the same time, the entity is required to deploy safeguards and measures to protect the rights and to implement data protection principles `by design and by default'~\cite[Art. 25(1)]{EU-Regulation679-2016}. While remaining solely responsible and accountable for meeting these requirements and for the processing as a whole, these entities will need the help of data analysis developers to understand and address the risks involved.\footnote{
    Further guidance was provided by the Art. 29 Working Party, replaced by the European Data Protection Board (EDPB).
}  This raises the questions of who is involved at which stages of data life cycles, how to analyse risks and employ safeguards? 
Additionally, data should be processed `lawfully, fairly and in a transparent manner', whereas processing should be `adequate and relevant and limited to what is necessary in relation to the purposes for which they are processed'.  This is the \emph{data minimisation} principle~\cite[Art. 5]{EU-Regulation679-2016}.  

When any other personal data processing activity is likely to result in high risks that an assessment of the risks and safeguards, a Data Protection Impact Assessment (DPIA) is required ~\cite[Art. 35]{EU-Regulation679-2016}. For the implementation and deployment of data processing activities, technology researchers and developers will need to take privacy and data protection rights and risks into account during the development process and incorporate necessary safeguards in order to preserve privacy~\cite[Recital 78]{EU-Regulation679-2016}.  Most Member States list biometric data processing as requiring such a DPIA.\footnote{
    See opinions by the EDPB on processing operations requiring a DPIA, \url{https://edpb.europa.eu/our-work-tools/consistency-findings/opinions_en}.}
The use of safeguards needs to be understood in a harmonized manner, when and how the privacy impact biometric data (in general, any sort of personal data) is too riskily in application deployment. Therefore, case studies---and moreover, a taxonomy on case studies---on how various forms of speech applications relate to another is compulsory. 
For communicating safeguards and their implementation between technology and legal communities, a common understanding (perhaps in the form of a taxonomy) is imperative.

\vspace{-0.25em}
\subsection{What are legal grounds to process sensitive data?}
\vspace{-0.25em}
Art. 9(1) of the GDPR~\cite{EU-Regulation679-2016} states that personal data falling into the category of sensitive data should not be processed, unless an exception applies as defined in  Art. 9(2) of the GDPR~\cite{EU-Regulation679-2016}. One of these exceptions relates to the processing of data that `are manifestly made public by the data subject'~\cite[Art. 9(2)(e)]{EU-Regulation679-2016}. The GDPR does not define this exception, but the Article 29 Working Party, an advisory body to the European Commission does in a non-binding opinion \cite{A29WP-LawEnforcementDirective-2017}. From that opinion, it is understood that data is manifestly made public when individuals \emph{deliberately} make their data public~\cite[p. 10]{A29WP-LawEnforcementDirective-2017}. This could mean that an individual who shares his/her own data via his/her personal web site could be considered to have expressed the intent to publicly disclose the data. An important distinction should be made between data placed into the public domain by others (e.g., information disclosed in a newspaper or broadcast on TV) and data voluntarily disclosed by the individual themselves.

From a legal perspective, it is not because the data is publicly available that it has been made available \emph{by the data subject}. Such a distinction might well apply to speech data; many audio/visual data can be found on the Internet , but it does not mean that the speech data contained in these files were made publicly available by the individuals to whom they belong. 

\vspace{-0.25em}
\subsection{Summary}
\vspace{-0.25em}
In order to interpret the impact of risks and regulation upon speech data, legal experts need a digestible DPIA. This is essential, since a risk assessment can only be performed if it is clear where the risk lies.  For example, if speech data is solely aimed at characterising (the biometric identity of) a speaker, all other information, for example race or emotion, may not be relevant in the processing of speech (in any form including raw, parametrised or otherwise human/automatic annotated speech).  Any divergence from such practice would not prevent possible `function creep', namely use of the same data for other purposes.  Safeguards to prevent such misuse should then be deployed by design.  
In order to understand and assess the risks of privacy intrusions, taxonomies are in high demand to facilitate the transparency and digestibility of concurrent speech research.

\section{Privacy, a Technical Perspective}
\label{sec:tech}
With a clearer picture of how privacy is characterised in terms of legislation, we explore here the potential for speech technologies to infringe upon privacy.  The treatment is intentionally high level---it is not intended to be exhaustive. The discussion is oriented around the legal reflections presented in Section~\ref{sec:legal}. With the aim of harmonising the work of the technical and legal communities, we elaborate on what speech communication is about (\emph{what is the focus of our research community?}). 

\vspace{-0.25em}
\subsection{What is speech [in] communication?}\vspace{-0.25em}
The Concise Oxford English Dictionary defines speech as \emph{the expression of or the ability to express thoughts and feelings by articulate sounds}. Communication is: \emph{1. the act of communicating. [\dots{}]; 2. the means of sending or receiving information, such as [by] telephone lines or computers; [\dots{}]}.  The verb `to communicate' means \emph{to share or exchange information or ideas. Convey an (emotion or feeling) in a non-verbal way. [\dots{}]}. Thus, in communications, speech is a medium to impart or exchange information. Speech is a particularly rich source of information, much of which is sensitive and personal.

\vspace{-0.25em}
\subsection{How is speech data captured, processed and stored?}\vspace{-0.25em}
The automatic treatment of speech data generally involves: (i)~speech capture and analog-to-digital conversion; (ii)~some form of frontend processing (e.g.\ acoustic feature extraction, subtraction of additive noise, deconvolution of convolutive noise, speech enhancement and segmentation of speech/non-speech intervals); (iii)~backend processing (e.g.\ projection of acoustic data to characteristic data representations, information modeling, classification, and system output calibration).

There is a potential for privacy intrusion from the very moment of speech data capture.  
Sophisticated algorithms have been designed to capture high-quality speech data from,
e.g., a single hand-held telephone microphone, multiple microphones used in modern smartphones (allowing for noise cancellation) and even microphone arrays (allowing for beamforming, the localisation and separation of a single, specific voice from a multi-speaker source).  
Technologies can bring substantial improvements to the quality of captured speech, while also resulting in the speech of a (distant) bystander being captured unwittingly.

\begin{figure}[!t]
    \centering
    \begin{tikzpicture}[font=\tiny]
        \node[anchor=south west,inner sep=0pt,outer sep=0pt] (image) at (0,0)     {\includegraphics[width=1\linewidth]{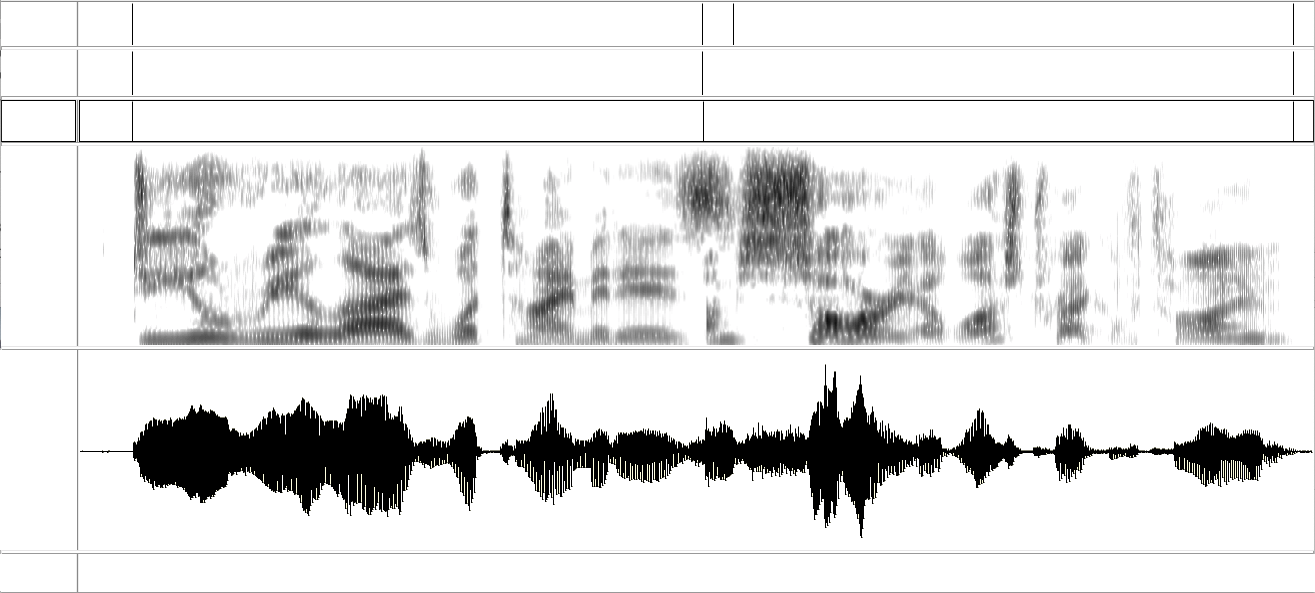}};
        \begin{scope}[x={(image.south east)},y={(image.north west)}]
            \node [anchor=north] at (0.244,0.085) {1\,s};
            \node [anchor=north] at (0.471,0.085) {2\,s};
            \node [anchor=north] at (0.698,0.085) {3\,s};
            \node [anchor=north] at (0.924,0.085) {4\,s};
            \node [anchor=north] at (0.03,0.15) {-8\,k};
            \node [anchor=north] at (0.03,0.420) {\hphantom{+}7\,k};
            \node  at (0.03,0.463) {1\,kHz};
            \node  at (0.03,0.508) {2\,kHz};
            \node  at (0.03,0.548) {3\,kHz};
            \node  at (0.03,0.590) {4\,kHz};
            \node  at (0.03,0.633) {5\,kHz};
            \node  at (0.03,0.673) {6\,kHz};
            \node  at (0.03,0.715) {7\,kHz};
            \node [align=center] at (.28,.795) {angry\vphantom{'y}};
            \node [align=center] at (.753,.795) {anxious\vphantom{'y}};
            \node [align=center] at (.28,.875) {Do you realise what time it is?\vphantom{'y}};
            \node [align=center] at (.753,.875) {I'm sorry, I lost track of time.\vphantom{'y}};
            \node [align=center] at (.28,.958) {female\vphantom{'y}};
            \node [align=center] at (.753,.958) {male\vphantom{'y}};
        \end{scope}
    \end{tikzpicture}
    \caption{Example of captured, processed and stored speech (sound extracted from \url{https://www.eslfast.com/robot/audio/dailylife/dailylife1901.mp3}).}
    \label{fig:annotation}\vspace{-0.5cm}
\end{figure}
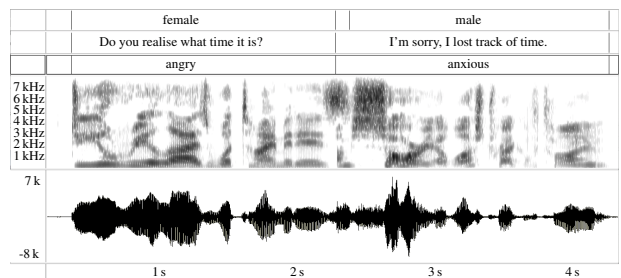

The form of frontend processing is usually adapted to the task, 
with the acoustic features derived from such processing forming the fundamental basis with which to suppress nuisance variation (noise), for example, but also as means of deriving subspace representations suitable for processing by backends.
Figure~\ref{fig:annotation} illustrates some of the information that backends can derive from speech data.  The bottom row shows a time domain representation and a  spectro-temporal representation, above.
It is most commonly from this spectro-temporal representation that the backend operates, deriving different sources and levels of information, many of which are potentially sensitive. 

\vspace{-0.25em}
\subsection{Why is speech data sensitive?}\vspace{-0.25em}
In the context of voice recordings with the aim to identify the speakers, without them knowing and without legal basis, it was decided that such action infringed the right to privacy \cite{ECHR-PGJHvsUK-2001}. Listening in or recording the content of telephone and other electronic conversations is generally forbidden under communication privacy, 
as further elaborated in more specific communication confidentiality provisions.  
Since a person's speech reflects their 
biological and behavioural characteristics, speech data is likely to qualify as \emph{sensitive data}. This becomes abundantly clear when realising that many of the 
attributes of speech signals, 
e.g.\ those derived by the frontend described above, have utility across a very broad of range of speech 
processing
operations. These operations focus on the processing of 
both 
verbal and non-verbal information, classifiable 
in terms of (i) linguistic, (ii) paralinguistic and (iii) extralinguistic information~\cite{LayerTruddgill-PhoneticLinguisticMarkersSpeech-CambrdigeUniPress-1979,Ephratt-LinugisticsParaExtra-JournalPragmatics-2011}. 

Behavioural influences stem from a person's geographical background, their social identity, ethnicity, socio-economic status and other learned phenomena such as personality, emotion, parental or familial influences and education. When combined, physiological/biological and behavioural influences are manifested as variations in the perceptual qualities of speech which can measured in terms of correlated physical quantities.  The latter are often expressed in terms of prosody (e.g.\ intonation, rhythm and stress), linguistic content (the words) and the spectral envelope (the timbre or `colour' of sound).

As illustrated in Figure~\ref{fig:annotation}, features derived by the frontend can be used by different backends different types of 
sensitive, personal information.  
Examples include the estimation of a speaker's age~\cite{Sadjadi-SpeakerAge-ICASSP-2016} and gender~\cite{Harb2005}, for instance.  The use of \emph{biometric data for uniquely identifying a natural person} can be the goal of speaker recognition (verification/identification technology)~\cite{Kinnunen-Li-Overview-Text-Independent-Speaker-Recognition-Speech-Communication-2010,HansenH15}.  Growing interests and studies in assisted living, ageing, medical diagnosis, emotion recognition and general well-being, e.g.~\cite{MencattiniMCTBBN14,VildaFBLMMMG09}, and a plethora of other health-related applications are clearly within the scope of \emph{data concerning health}.  That speech signals can also be used to characterise ethnicity \cite{Hanani-AccentEthicsRecognitionBE-CSL-2013} indicates that speech data could fall under the definition of \emph{personal data revealing racial or ethnic origin}.  If a machine can decipher spoken language, then there is no reason why it could not also predict from the annotation of the spoken words (and conversations) that person's \emph{political opinions, religious or philosophical beliefs}.  Lastly, some research claims that sexual preferences can also be predicted through speech data~\cite{Gaudio94}. Speech data could also fall within the scope of \emph{data concerning a natural person's sex life or sexual orientation}.

\vspace{-0.25em}
\subsection{What safeguards are there?}\vspace{-0.25em}
\label{sec:safeguards}

A recent survey of privacy preserving safeguards for speech data is presented in~\cite{Nautsch-PreservingPrivacySpeech-CSL-2019}.
A number of different techniques have emerged in recent times.  
Homomorphic encryption (HE) \cite{pathak2013,Nautsch-2CovHE-Odyssey-2018} is a form of cryptosystem designed to process speech data in the encrypted domain,
though alternative data representations are generally required since speech data is typically stored as floating-point data, whereas cryptosystems operate on integer data. 
Garbled circuits~\cite{portelo2014gc} involve the splitting of data into randomised components, each of which is then processed by independent servers that jointly and securely compute an operation upon speech data without privacy leakage.  
Cancelable biometrics \cite{Billeb2015IET,mtibaa2018cancelable} are based on model binarization, which also find use in hashing techniques~\cite{jimenez2015smh}, irreversible but comparable speech representations. 
Differential privacy techniques~\cite{shokri2015privacy} preserve privacy by learning data representations from which information not relevant to a given application is suppressed (e.g., information on a speaker's identity is removed from a representation used in speech recognition).
Finally, hardware-assisted techniques (e.g., based on the Intel SGX architecture)~\cite{brasser2018voiceguard} can complement software-based techniques.
Most of these techniques can be deployed as `addons' to deliver privacy preservation in the case of otherwise unprotected systems.  Alternatively and preferably, they can be incorporated from the moment of system conception according to the 
`privacy by design' principle. 
Somewhat orthogonal approaches to privacy preservation include identity obfuscation/speaker de-identification \cite{Jin-SpeakerDeIdentification-ICASSP-2009,Bahmaninezhad-SpeakerDeiIdentification-Odyssey-2018} (for speech appearing to be of another).

\section{On the Need for Taxonomies}
\label{sec:taxonomoies}

Clearly, the legal and technical communities lack a shared understanding of the implications of the GDPR as regards speech data.  The following proposes the anchors for classification schemes with semantic relationships (taxonomies) to facilitate the discussion that will be needed to establish an initial level of harmonisation.  While currently lacking, it is crucial to the preparation of DPIAs and future dialogues between the legal and technology communities that advances in technology are accompanied with adequate provisions for privacy preservation.

 


%
\textbf{1) Information in speech merits protection:}
%
It is first necessary to distinguish between the different types of information demanding protection.
Such taxonomy classes could compare sensitive, personal, (legally) non-personal but protection-worthy, and unprotectable data derivable from speech. To define these and their relations in a digestible manner for non-experts, communication models (e.g.,~\cite{schulzvonThun-4sidesModel-1981}) may be be useful tools.


\textbf{2) Capture of speech signals:} The manner in which speech is captured (single/multiple microphones), in addition to sensor configurations and locations (distance from speakers, location, single/multi-room) influences potential privacy intrusions (the number of persons from whom speech is captured).
Class relations could emphasise on unwittingly or consensually captured speech (on own devices or of others).

%

\textbf{3) Processing of speech data:} Clear, understandable descriptions of the purpose and interrelations between research areas are needed so that the legal community is able to form legislation with a view as to how it will impact upon speech technology and privacy in speech data.  
The editors information classification scheme (EDICS) may serve useful here.




\textbf{4) Storage of speech data:} Some level of transparency is required concerning the means (e.g.~as raw data or other representations) and location of speech data storage (e.g.~strictly on a user's mobile device and in the cloud; de/centralised), in addition to access policies.  Clarity will be vital to the legal community so as to identify data processors and controllers, and to determine the potential for data to
leave the EU. 

\textbf{5) Entities in speech data lifecycles:} \emph{Who (i)~creates, (ii)~integrates, (iii)~operates, (iv)~provides and (v)~owns (sub-)system components?} Since modern speech processing systems typically run on multi-party server infrastructures, transparency is necessary for conducting DPIAs that outline safeguards.

\textbf{6) Case studies:} 
As a means of managing the almost limitless variability in speech data applications (in e.g., smart homes, health care, social media, eLearning platforms), taxonomy classes for use cases need defining in order to facilitate the dialogue between legal and technical communities.  Class relations might be based on if senders/recipients in communication are peers and how information flows in their communication. Only then can the requirements for safeguards be determined.


\textbf{7) Technology safeguards:} Safeguards such as encryption, should be designed according to the specific use case and DPIA. 
Safeguards can either enhance existing technology, i.e.~\emph{privacy as an addon}, or as \emph{privacy by design} principles and also be used for de-identification/doxxing.
Solutions can be classified according to the attributes of the underlying techniques, e.g., cryptographic technologies, security proofs, resource demands and assumptions. 
Cryptographic technology is needed that facilitates the (real-time) demands of speech technology; applying conventional encryption on waveforms is likely to render any inference in speech processing computationally useless.

Even so, to satisfy strict DPIA interpretations, a legal perspective might demand the obfuscation/segregation of features that could potentially describe sensitive data which is not relevant to the use in a certain speech application, e.g., the use of soft-biometric information such as ethnicity is not ultimately necessary for speaker recognition.  From a technical perspective, however, it may not be possible to meet these demands with current capabilities, because features that reveal sensitive data are derivable from many levels, e.g., even if acoustic features that indicate certain accents could be segregated (inducing artefacts lowering intelligibility), the textual representation of uttered speech might still comprise linguistic features that reveal the geographical background of the speaker.

\section{Conclusions}
\label{sec:conclusion}

\added{This paper summarizes the first reflections of legal and technology communities upon the GDPR and speech data.}
For non-experts, a grasp of speech as a data modality can be as challenging as achieving harmonisation between two communities (much more so than for fingerprint or facial data). Here, harmonising legal and speech research is challenging.
The speech community must understand 
the legal perspective regarding privacy legislation just as the 
legal community must understand the technological implications. This common understanding will only be achieved by reaching out to our colleagues and by collaborating on the preparation of policy papers (opinions); policy papers that will eventually lead us to (better informed) legislation, and better designed products and services.

Provision and interpretation of the law, such as for the implementation of privacy safeguards, need to make technology-agnostic sense.  In outlining some reflections of the legal and technology communities, this contribution and proposed taxonomies is a first step in this direction.  While it focuses on the implications of the GDPR, this is certainly not the only legislation relevant to privacy in speech data.  Even so, the proposed taxonomies should also be relevant to privacy legislation outside of Europe. 
Future work should develop privacy safeguards that encompass not only the protection of speech data observations and representations, but safeguards that are appropriate and that account for the nature of speech as a communication medium.
Clearly, though, the dialogue between our two communities must continue and are in all of our interests.

\smallskip\textbf{Acknowledgements.}
The authors thank Bhiksha Raj for his feedback in the beginning of our discourse. This work is partially funded by: the Horizon 2020 research project PDP4E (contract number 787034); FCT (reference UID/CEC/50021/2019); ANR projects Voice Personae and RESPECT, and Omilia -- Conversational Intelligence.

\vfill
\balance

\bibliographystyle{IEEEtran}
\bibliography{main}

\end{document}